# Development of A Load Control Algorithm to Enhance Energy Sustainability for the International Space Station

Ethan Herwald, Garrett Holliday, and Murat Kuzlu
*Engineering Technology, Old Dominion University, VA, USA*

Onur Elma
*Electrical Engineering, Yildiz Technical University, Istanbul, TURKEY*

Umit Cali
*Electric Power Engineering, Norwegian University of Science and Technology, Trondheim, Norway*

*Abstract*— This paper presents a load control algorithm for control of energy sources and loads to enhance energy sustainability and reliability of the International Space Station (ISS), which is a large spacecraft in orbit around Earth. In this paper, the ISS electric power system was simulated in MATLAB/Simulink to be able to evaluate the performance of the developed algorithm in a simulated environment. This study also aims to emphasize the importance of load control algorithms on energy sustainability for critical systems, like ISS, having limited energy sources.

## I. Introduction

The international space station (ISS), launched in 1998, is a complicated and large-scale system, which provides a scientific laboratory with the ability to conduct experiments in space. The ISS has a complex structure, consisting of various components, such as logistics carriers, radiators, solar arrays [1]. The energy required for the operation of the ISS is present, just challenging to harvest. To aid in energy harvesting, NASA has eight solar arrays deployed, which produce 84 to 120 kW on average during the insolation period and eclipse period of the orbit.

Due to their inherent remoteness and limited energy resources in a space environment, all space applications require the efficient distribution of the stored electrical energy. The ISS electrical power system is the largest space-based power system and has driven advanced technologies with operational challenges [2]. Furthermore, the ISS is not always in direct sunlight due to orbiting the Earth. This means the energy generation is limited during non-sunlight periods. The space station relies on the energy storage system, consisting of rechargeable battery modules, onboard sensors, and control components to provide continuous energy. Energy management is crucial to support the operation of the ISS continuously and efficiently due to the nature of the space environment. The ISS needs constant electrical energy to support all space operations [3]. The system of solar arrays, made of thousands of cells, is the only practical way to generate electrical energy [4]. Real-time monitoring and control of the solar arrays are also important to maximize power production during the ISS operation. The storage system, i.e., batteries, is used to store excess solar array energy during periods of sunlight and provide power during periods when the station is in Earth's shadow [5]. During the insolation part of the orbit, the batteries are recharged. The generation, distribution, and storage of the system need to be as efficient as possible to operate the station and sustain life support continuously for the ISS.

The most important aspect of space-based applications is the development of a reliable microgrid. A microgrid is a self-sustaining electrical system composed of a generation process, energy storage system, and an integrated load control network used to ensure the efficient operation of the grid. Obviously, this can be a challenging task when designing a reliable microgrid for space applications. In the event that the designed microgrid will be aboard a manned spacecraft, there will need to be enough electrical energy generated to compensate for the everyday needs of the crew. The ISS is also an example of a microgrid, which is a local energy grid with control capability [6]. Creating a model that simulates the generation, storage, and distribution of the ISS's power system is an important step to becoming more familiar with microgrid technology for space-based applications [7].

In the study [8], the authors analyzed the common space solar power generation, management, and distribution method, as well as proposed a new hybrid power system structure for space solar arrays, which is constructed and a modular multi-converter with a serial-parallel combination control strategy is investigated. The authors in [9] provided details of the ISS electrical power system architecture, including power generation, energy storage, distribution, and potential future space energy generation and storage technology development. The study [10] provided a comprehensive work on ISS electrical system performance and operational lessons learned, covering details of the electrical power system architecture and its performance. The study [11] presented details of the ISS energy storage system with the on-orbit operation and alternative energy storage system design for high reliability and long service life. It is indicated that the flywheel energy storage system (FESS) is a good candidate, which offers more energy, higher efficiency, and three times the operating life of the ISS energy storage system, i.e., nickel-hydrogen battery (Ni-H2). Ni-H2 batteries were used to store ISS electrical energy. However, these batteries had approached their end of useful life, and the ISS Program began the development of Lithium-Ion (Li-Ion) batteries to replace the Ni-H2 batteries in 2010 [12]. In the study [13], the authors analyzed the small-signal stability for the Japanese Experiment Module (JEM) of ISS, which can be expanded for all the space electric power systems.

In this study, the ISS electrical power system model of the U.S. segment is simulated in MATLAB/Simulink in order to apply load control algorithms to enhance energy sustainability for the ISS operations. A load control algorithm is developed for the efficient distribution of stored electrical energy and to handle various scenarios. A case study is conducted on the model to determine the strengths and weaknesses of the load control algorithm under various generation scenarios.

## II. ISS Electrical Power Distribution Architecture

The electrical power system (EPS) of the international space station is critical to sustain and operate ISS operations with limited energy resources. These operations also include water dumps, visiting spacecraft, and vehicular activities. In addition, the demands of the new science and crew modules cause more energy requirements for international space stations [14]. For these reasons, the electrical power system of the ISS will be investigated to find the best way to enhance energy sustainability by developing various energy management approaches. Fig. 1 depicts the ISS electrical power system architecture.

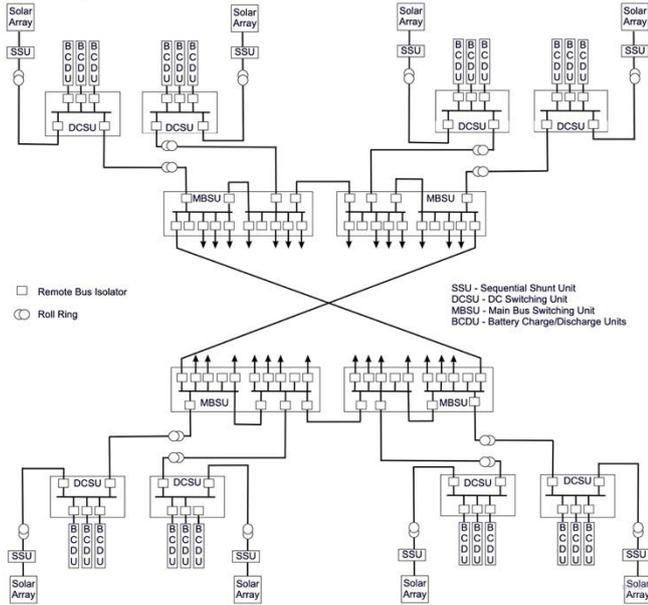

Fig. 1. The overview of ISS electrical power system [15]

The ISS has a very complex electrical power architecture with five subcomponents. The first subcomponent is power generation, which consists of the eight solar array wings. The second subcomponent is the regulator, which consists of the sequential shunt unit (SSU) and the DC switching unit (DCSU). The regulator subcomponents route the generated power to the battery control, which is the third subcomponent. The battery control sub-component consists of the battery charge-discharge unit (BCDU) and batteries. The fourth subcomponent is the DC power distribution. The DC distribution system on the ISS creates challenges on how to analyze the system. The last subcomponent is electrical loads throughout the station, which is drawing power from an array at any given moment, constitutes a specific level of power demand.

The generated electrical power by solar PV is sent to the SSU and then to DCSU. The voltage from the arrays is regulated by a device called the SSU. All the arrays are attached to the station through pivotal joints. These pivotal joints, named beta and alpha gimbals, allow the axis for rotation, so the solar arrays are continually oriented in the most optimal direction. The SSU transfers the regulated voltage through the beta gimbal to the DCSU. While the ISS is in the insolation portion of the orbit, the DCSU allows power to flow to the BCDU. During the dark or eclipse portion of the orbit, the DCSU receives power from the batteries through the BCDU, which regulates this power. From the DCSU, the power is fed into the main bus switching unit (MBSU). The MBSU is essentially a complex network of relays. From the MBSU, the power is either fed to the American Russian conversion unit or DC-DC converters. The DC-DC converters convert the primary voltage level of 160 volts to the secondary voltage level of 120 volts [16]. The U.S. portion operates at a voltage of 120 $V_{dc}$, and the Russian portion operates at both 28 $V_{dc}$ and 120 $V_{dc}$. The U.S. portion operates independently from the Russian segment. DC to DC converters tie the two portions together to allow for power-sharing under certain circumstances [17]. However, the Russian segment of the ISS does depend on the American generation in numerous cases. For the Russian segment to receive power for the U.S. segment, another converter is used. This converter is known as the American/Russian converter unit (ARCU). The energy produced was originally stored in 38 lightweight nickel-hydrogen batteries packaged in an orbital replacement unit (ORU). Two ORU modules make up one battery, and there were 24 batteries on the International Space Station. The nickel-hydrogen batteries have since been replaced with more advanced lithium-ion batteries.

## III. The Proposed Intelligent Algorithm

The objective of the developed algorithm is to recommend possible operation and control strategies under a varying load demand. This process is illustrated in Fig. 2. As shown, the ISS energy management controller takes the following inputs: (1) Solar arrays (power generation - kW), (2) Channel/load priority (L1>L2>L3…), and (3) Batteries (energy storage capacity - SoC - %). Next, the algorithm decides to toggle loads based on these three inputs. The output of the controller is a set of possible energy usage targets that can be achieved by managing the internal resources and loads. It aims to present the ability to perform energy management control at the load level by their type and priority during the high peak and low battery charge. With this ability, the highest priority loads can be secured, and the non-critical loads can be shed to match the limitations in supply.

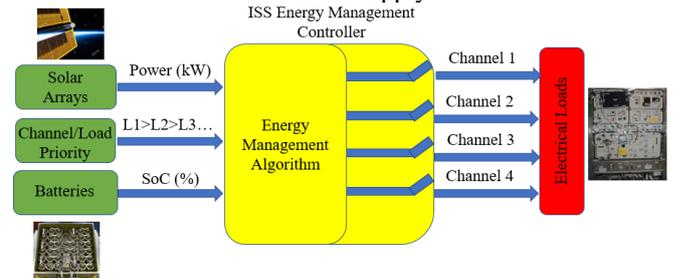

Fig. 2. The block diagram of the developed ISS energy management algorithm

## IV. Load Modeling

The ISS load bank is broken into four main channels. Each of these channels consists of critical and non-critical loads. The critical loads are the battery units, atmosphere control, crew system, control system, communications, the main computer, and the air pumps. Table I gives a comprehensive list of the loads in the ISS and a breakdown of the four different power channels.

The United States portion of the International Space Station, under normal operating conditions, will deliver an average of 76,000 Watts. Including the Russian segment, the ISS delivers nearly 105 kilowatts.

TABLE I. THE LIST OF ISS LOADS [18]

| Channel 1 | Power (kW) | Channel 2 | Power (kW) |
|---|---|---|---|
| Battery Unit | 6.645 | Battery Unit | 6.645 |
| Fan | 1.605 | Fan | 1.605 |
| Atmosphere Controller | 1.2 | Atmosphere Controller | 1.2 |
| Crew System | 0.575 | Crew System | 0.575 |
| Control System | 0.82 | Control System | 0.82 |
| Communications | 0.47 | Communications | 0.47 |
| Lighting Bank | 1.08 | Lighting Bank | 1.08 |
| Main Computer | 0.385 | Main Computer | 0.385 |
| Robotic Workstation | 0.895 | Robotic Workstation | 0.895 |
| Robotic Arm | 3.21 | Robotic Arm | 3.21 |
| Air Pump | 1.15 | Air Pump | 1.15 |
| **Total** | **18.035** | | **18.035** |
| Channel 3 | Power (kW) | Channel 4 | Power (kW) |
| Battery Unit | 6.645 | Battery Unit | 6.645 |
| Fan | 0.535 | Fan | 1.07 |
| Lighting Bank | 0.72 | Lighting Bank | 0.36 |
| Experiment U.S. 1 | 4.25 | Experiment U.S. 2 | 3.005 |
| Experiment U.S. 3 | 2.275 | Experiment U.S. 4 | 2.26 |
| Experiment Russian 1 | 2.715 | Experiment Russian 2 | 3.2 |
| Experiment Russian 3 | 1.845 | Experiment Japan 2 | 0.92 |
| Experiment Japan 1 | 1.985 | Experiment Japan 3 | 3.46 |
| **Total** | **20.97** | **Total** | **20.92** |

The total ISS load is modeled by creating four (4) subsystems labeled channels 1-4, as shown in Fig. 3. Each subsystem is connected to a scope individually so that the power consumption of each channel can be monitored. In addition, the combined load amount can be monitored with a sum function. The input to each channel subsystem is a decimal input that comes from the load control algorithm and is converted to binary within the subsystem.

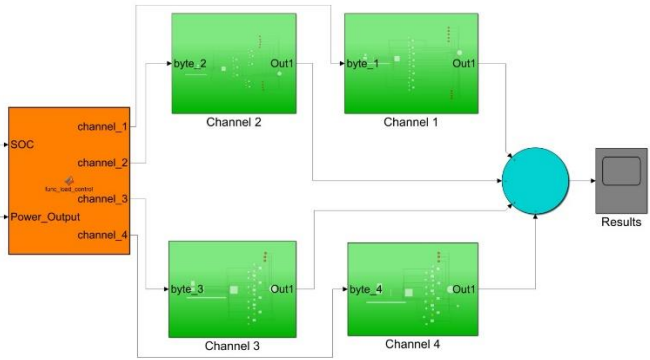

Fig. 3. ISS load model with four (4) subsystems

Once converted, the binary resultant is sent to a demultiplexer, directly controlling the switches that toggle the loads. The switches are used as relays. When the switch is cleared, the connection is made to the load. When the switch is set, the load is turned off. Another sum function is added to sum all of the load's power consumption. This design allows us to control individually selected nonessential loads by sending one decimal number from the load control function to each channel. The essential loads are represented by a constant that cannot be turned off.

## V. CASE STUDY DESCRIPTION

The case study uses a developed load control algorithm and tests the algorithm response to various generation scenarios. The essential operation of the load control algorithm is to perform load shifting. In order to accomplish load shifting, a load control algorithm was developed. The algorithm monitors the battery state-of-charge (SoC) and adjusts the loads accordingly. It is important to note that each channel has loads that are considered essential for the ISS to operate, i.e., battery unit, atmosphere controller, crew system, control system, main computer, air pump. Therefore, these loads cannot be toggled off and must run under all state-of-charge conditions. The total estimated power consumption on channels 1 and 2 is given in Eq. (1)

$$P_{1\&2} = 2*(6.645+1.2+0.575+0.82+0.47+0.385+1.15) = 22.49 \text{kW} \quad (1)$$

The essential loads for channels 3 and 4 are the six battery units. The net power that will be consumed during all SoC conditions for channels 3 and 4 is shown in Eq. (2):

$$P_{3\&4} = 2*(6.645) = 13.290 \text{kW} \quad (2)$$

Therefore, it is concluded that under all SoC conditions, the power demand of the loads is shown in Eq. (3):

$$P_{essential} = 22.49 + 13.290 = 35.78 \text{kW} \quad (3)$$

The total power demand under normal circumstances can be calculated as in Eq. (4):

$$P_{total} = P_1 + P_2 + P_3 + P_4 = 18.035 + 18.035 + 20.97 + 20.92 = 77.96 \text{kW} \quad (4)$$

The load control algorithm monitors the SoC of the batteries. Arbitrarily, if the SoC is less than 40%, all nonessential loads will be turned off. Therefore, as previously calculated, the load demand will be 35,780 watts. In order to achieve this condition, a value of 255 is sent to each channel. Once this value arrives at the channels, it is converted to binary and toggles the switches corresponding to the individual loads. It is important to mention that a binary bit value of 1 will toggle the switch off. Therefore, 255 will result in all nonessential loads toggling off.

In the event that the SoC is less than 80% but greater than 40%, the certain nonessential load will be toggled off. These nonessential loads were determined to be the robotic workstations, Canadian robot arm, the U.S. experiments 1 and 4, Russian experiment 3, and Japan experiment 1 and 2. These loads were selected because they consumed the most power per channel and are not essential for ISS operation.

The total power consumption of these loads is calculated in Eq. (5):

$$P_{estimated} = 2*0.895 + 2*3.21 + 4.25 + 1.845 + 1.985 + 2.26 + 3.46 = 22.010 \text{kW} \quad (5)$$

With these loads toggled off, the power demand will be shown in Eq. (6):

$$P_{estimated} = 77.96 - 22.010 = 55.95 \text{kW} \quad (6)$$

Once the batteries reach a SoC greater than or equal to 80%, it is determined safe to toggle all loads on. This will result in power consumption of 77,960 watts. According to Table I, essential loads for channels 1 and 2 are the six battery units, two-atmosphere controllers, two crew systems, two communication modules, two main computers, and two air pumps.

In order to test and observe the effects of the load control diagram, a generation scenario function was created. Within this function, the electrical current generated by the solar panels is altered based on the orbit minute. The power is calculated using the voltage and current of the panel. Since only the current changes, the power will decrease by the same factor as the current, as shown in Fig. 4.

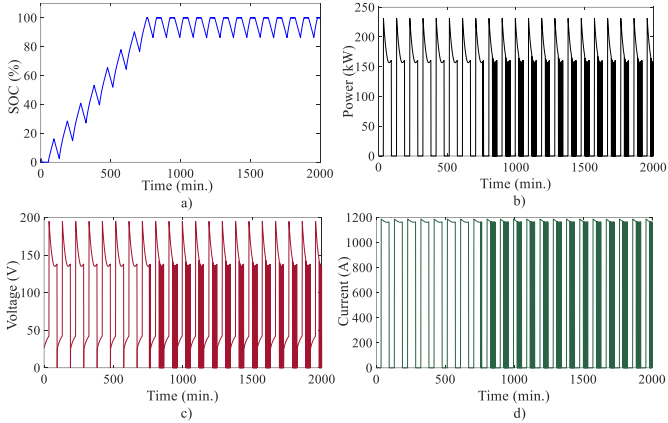

Fig. 4. ISS simulated SoC, PV power, PV voltage, and PV current under ideal circumstances

This scenario is used as a basis and shows the battery SoC climbing for 830 minutes (9.02 orbits) until it reaches 100% under the ideal generation. At this point in the simulation, it does not allow to charge the batteries from the solar arrays to prevent the overcharging, i.e., the delivered PV power to the batteries is zero. It also affects the PV voltage and current. For the base case, the load control algorithm is not integrated into the system, and batteries deliver the full demand of the ISS without any load control.

In order to visualize the effects of a catastrophic failure of the generation aboard the ISS, a scenario was developed that decreases the generation by 10% every 200 minutes of orbit. The gradual decrease in generation reaches a steady state after 1000 minutes of orbit with a total reduction in generation of 50%. The effects of this scenario are witnessed in Fig. 5.

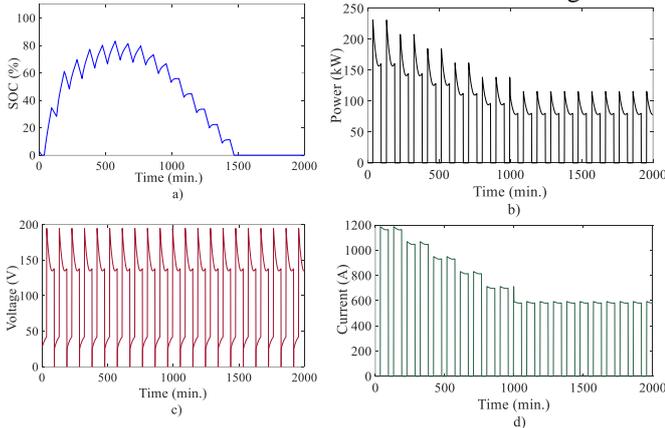

Fig. 5. ISS simulated SoC, PV power, PV voltage, and PV current without the load control during a catastrophic generation failure

As can be seen in Fig. 5, the SoC peaks around 50% and immediately begins to decrease as a result of the decrease in the generation due to a catastrophic generation failure.

## VI. RESULTS AND DISCUSSION

In order to show the effect of the load control algorithm, a basis simulation was conducted to show the duration until full SoC is achieved while controlling the loads. It is essential to investigate the impacts of the load control algorithm on the SoC during various generation scenarios. The first scenario is identical to the catastrophic generation failure depicted in Figure 5. The effect of the load control algorithm can be seen in Fig. 6.

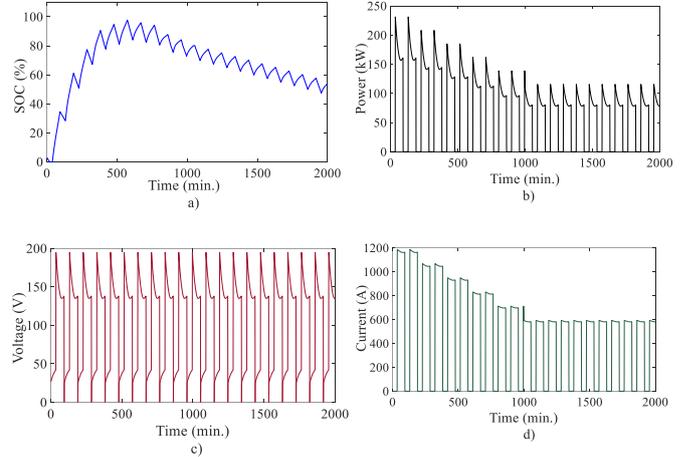

Fig. 6. ISS simulated SoC, PV power, PV voltage, and PV current with the load control during catastrophic generation failure

Fig. 6 shows the results of introducing the load control algorithm during a catastrophic generation failure event. During this simulation, when the SoC reaches 80%, all loads of the ISS are turned on and consuming 77,960 watts from the batteries. Once the SoC falls below 80% but remains above 40%, the certain nonessential loads mentioned in the previous section are disconnected. When the SoC falls below 40%, all nonessential loads are turned off to minimize the effects of the failed generation. With the generation being reduced by 50%, the SoC begins to decrease until it reaches 40%. At this point in the simulation, all nonessential loads are disconnected, which results in a stabilized SoC of around 40%.

Comparing Fig. 5 with Fig. 6, it is concluded that the load control algorithm fulfills its purpose by regulating the SoC under varying generation conditions. By introducing the algorithm, the ISS is still able to operate until the generation failure is resolved. This could reduce the risks of the space crafts' personnel by giving them a chance to proceed the needed repair the failed generation on time.

A final scenario was simulated to determine the breaking point of the load control algorithm. In this scenario, the generation is again reduced by 10% every 200 minutes; however, once the simulation time reaches 1200 minutes, the generation is drastically reduced by 70%. The results of this simulation can be seen in Fig. 7 shows the power consumption during the simulated interval.

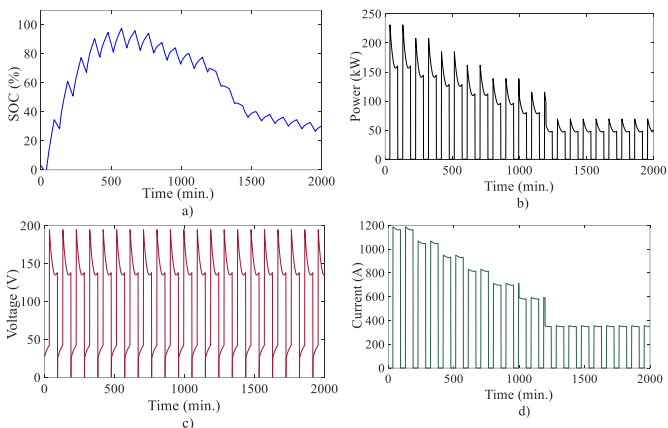
Fig. 7. ISS simulated SoC, PV power, PV voltage, and PV current with the load control during 30% generation capacity

The final conducted simulations are demonstrated in Fig. 7 where the full impact of the algorithm on the SoC of the batteries is observed. During the first 275 minutes of the simulation, the SoC gradually increases until it reaches 100%. When the SoC reaches 40% and 80%, different power consumptions occur. This full power consumption lasts until approximately 830 minutes of orbit. At this point, the consumption dominates the generation, and the load control algorithm turns off certain nonessential loads. The oscillation of power consumption is a result of the SoC periodically exceeding the thresholds allowing for different consumption rates. After 1200 minutes of orbit, the generation is dramatically reduced by 70% resulting in the SoC plummeting. Once the SoC reaches 40% at approximately 1400 minutes, the load control algorithm disconnects all nonessential loads. A result of this disconnection is a slighter decrease in SoC. Since the SoC is still decreasing and all nonessential loads are turned off, the ISS will not be able to operate under this condition.

## VII. CONCLUSION

For a microgrid to be self-sustaining, the ability to control the power consumption of the loads for maximum efficiency is mandatory. Without this ability, the discharge rate of the storage device will not be controlled. This could ultimately result in loss of power to the system. In this study, the ISS electrical power system model is simulated in MATLAB/Simulink. A load control algorithm is developed for the efficient distribution of stored electrical energy and to handle various scenarios. The developed model includes an accurate representation of the loads aboard the ISS. To feed power to the loads, a generation model of the ISS was created. To control the rate of power consumption at loads, a load control diagram was developed. Finally, to test this load control algorithm, a function dedicated to altering generation values was implemented to observe the reaction of the system. It can be concluded that the load control algorithm fulfills its purpose in conserving the SoC of the batteries of the ISS. The results indicate that the SoC may plummet to 0% without the algorithm during a catastrophic generation failure, while the SoC is maintained at around 40% with the algorithm.

Although this model represents the ISS power system, with some slight adjustments, it could be used to simulate the power system of any microgrid.

ACKNOWLEDGMENT

This work was supported in part by the Virginia Space Grant Consortium (VSGC) through the 2019 New Investigator Program (NIP).

REFERENCES

[1] Glenn Research Center, Powering the Future, [Online]. Available: https://www.nasa.gov/centers/glenn/about/fs06grc.html.
[2] National Aeronautics and Space Administration (NASA), The Electric Power System of the International Space Station A Platform for Power Technology Development, March 2010.
[3] EDN Network, International Space Station (ISS) power system, [Online]. Available: https://www.edn.com/design/power-management/4427522/International- Space-Station-- ISS--power-system.
[4] M. Kuzlu, et al. "Modelling and Simulation of the International Space Station (ISS) Electrical Power System." International Transactions on Electrical Energy Systems, 2021, doi: 10.1002/2050-7038.12980.
[5] National Aeronautics and Space Administration (NASA), [Online]. Available: https://www.nasa.gov/mission_pages/station/structure/elements/solar_arrays-about.html.
[6] Department of Energy (DoE), How Microgrids Work, https://www.energy.gov/articles/how-microgrids-work
[7] National Aeronautics and Space Administration, Lessons Learned From Developing Advanced Space Power Systems Applied to the Implementation of DC Terrestrial Micro-Grids, [Online]. Available: https://ntrs.nasa.gov/archive/nasa/casi.ntrs.nasa.gov/20150010351.pdf
[8] L. Wang, D. Zhang, J. Duan and J. Li, "Design and Research of High Voltage Power Conversion System for Space Solar Power Station," 2018 IEEE International Power Electronics and Application Conference and Exposition (PEAC), Shenzhen, 2018, pp. 1-5.
[9] E. B. Gietl, E. W. Gholdston, B. A. Manners and R. A. Delventhal, "The electric power system of the International Space Station-a platform for power technology development," 2000 IEEE Aerospace Conference. Proceedings (Cat. No.00TH8484), Big Sky, MT, USA, 2000, pp. 47-54 vol.4.
[10] M. Savoy and T. Miller, "International space station electrical power system performance and operational lessons learned," IECEC '02. 2002 37th Intersociety Energy Conversion Engineering Conference, 2002., Washington, WA, USA, 2002, pp. 93 doi: 10.1109/IECEC.2002.1391985.
[11] H. Oman, "International Space Station power storage upgrade planned," in IEEE Aerospace and Electronic Systems Magazine, vol. 18, no. 5, pp. 32-39, May 2003, doi: 10.1109/MAES.2003.1201457.
[12] E. Schwanbeck and P. Dalton, "International Space Station Lithium-ion Batteries for Primary Electric Power System," 2019 European Space Power Conference (ESPC), Juan-les-Pins, France, 2019, pp. 1-1.
[13] Masaaki Komatsu and Satoru Yanabu, "Analysis of the small signal stability for the international space station/JEM electric power systems," 2008 IEEE 2nd International Power and Energy Conference, Johor Bahru, 2008, pp. 106-111, doi: 10.1109/PECON.2008.4762454.
[14] Reddy, S. Y., Iatauro, M. J., Kurkl, E., Boyce, M. E.; Frank, J. D.; and Jonsson, A. K., Planning and monitoring solar array operations on the ISS. In Proc. Scheduling and Planning App. Workshop (SPARK), ICAPS, 2008.
[15] Benjamin Loop, Simulation Environment for Power Management and Distribution Development, EnergyTech2015, Nov 30, 2015 - Dec 02, 2015. Cleveland, Ohio.
[16] Huckins, E., Ahlf, P. (1994). Space Station power requirements and issues. IEEE Aerospace and Electronic Systems Magazine, 9(12), 3-7.
[17] Gietl, E. B., Gholdston, E. W., Manners, B. A., Delventhal, R. A. (2000, March). The electric power system of the international space station-a platform for power technology development. In 2000 IEEE Aerospace Conference. Proceedings (Cat. No. 00TH8484) (Vol. 4, pp. 47-54). IEEE.
[18] National Aeronautics and Space Administration. (n.d.) Solar Energy for Space Exploration Teacher's Guide. [Online]. Available: https://www.nasa.gov/pdf/160491main_SESETeachersGuide_dc3.pdf.